\documentstyle[]{l-aa}

\def\msun{M$_\odot$}
\def\approxlt{\rlap{$<$}{}_{{}_{{}_{\textstyle\sim}}}}
\def\approxgt{\rlap{$>$}{}_{{}_{{}_{\textstyle\sim}}}}
\def\about{$\sim$}
\begin{document}
\title{Search for neutron star spin periods in X-ray bursts}
\author{H. C.\, Jongert
   \and M. van~der~Klis}
\offprints{M. van der Klis, Astronomical Institute ``Anton Pannekoek''}

\institute{Astronomical Institute `Anton Pannekoek' and Center for High Energy
Astrophysics, Kruislaan 403, NL-1098 SJ Amsterdam, The Netherlands}

\thesaurus{08(08.14.1;08.15.1;13.25.1)}

\date{Received ; accepted }

\maketitle

\begin{abstract}

We report a search for pulsations in 147 Type~1 X-ray bursts observed with
EXOSAT in 10 X-ray burst sources. Instead of treating each burst separately,
we incoherently averaged the power spectra of all of the bursts of a given
burster to improve statistics. No periodicities were detected; 99\% confidence
upper limits on the modulation depths of possible periodic signals with
frequencies between 1 and 2048~Hz range from 2 to 20\% depending on source and
frequency range.

\end{abstract}

\section{Introduction}

Searches for the predicted millisecond pulsations in the persistent X-ray flux
of low-magnetic-field neutron stars have, so far, had no success (see Vaughan
et al. 1994 and references therein). Unlike the persistent X-ray emission, the
result of the release of gravitational energy in the accretion process, X-ray
bursts arise through nuclear combustion of accreted matter on the surface of
the neutron star (see Lewin et al. 1993 for a review). The nuclear energy
generation does not necessarily occur uniformly over the neutron star surface.
According to theory (Fryxell and Woosley 1982, Nozakura et al. 1984, Bildsten
1995), the accumulated nuclear fuel first ignites at the point on the neutron
star surface where it reaches the critical ignition column density, and then
spreads by a process of convective combustion to all adjacent areas on the
surface where the column density is larger than the critical propagation
column density. Non-homogeneous accretion, or geometrically irregular steady
burning will lead to areas with super-critical ignition column densities with
adjacent areas with super-critical propagation column densities bounded by
areas where the propagation column density is sub-critical (see also Chau et
al. 1995).  This will lead to X-ray burst emission that is non-uniform over
the neutron star surface, causing periodic variability in the burst flux as
the star spins.  According to Bildsten (1995), the formation of isolated fuel
supplies is a process that is strongly influenced by the accretion rate. At
low accretion rate nuclear burning in the bursts is expected to occur
uniformly over the surface, and at high accretion there are no bursts as all
fuel is consumed in persistent nuclear burning. At intermediate accretion
rates the above described ``patchy'' burning process, which would make the
neutron star spin period visible, could take place.

Periodicity searches in X-ray bursts have been previously performed by Mason
et al. (1980), Sadeh et al. (1982), Murakami et al. (1987), and Schoelkopf and
Kelly (1991), but so far none of the reported periods has been confirmed. The
most convincing detection was that by Schoelkopf and Kelly, who found a 10\%
half-amplitude, $4.1\sigma$ periodicity at 7.6~Hz in a burst from the soft
X-ray transient Aql~X-1, observed with the Einstein Monitor Proportional
Counter.

In the present work, we make use of bursts observed with the EXOSAT
observatory in some X-ray burst sources to perform a burst periodicity search
of increased sensitivity. We calculate the power spectra of all bursts of a
given burster and average them. To our knowledge this method has not been
previously applied.

\section{Data}

We used data obtained with the 2--20~keV EXOSAT ME argon detectors (Turner et
al.\ \cite{Turner:EXOSAT}, White and Peacock 1988). Data were recorded using a
variety of high time resolution modes and at resolutions of 0.25 to 32
milliseconds. When possible, the 5--35~keV xenon detectors, which record
mostly background counts, and data from detectors not pointed to the source
were excluded from the analysis. The count rates during the bursts were of
course strongly variable; peak count rates varied between 100 (1705-44) and
$10^{4}$ (1820-303) counts/second.

We selected a total of 147 bursts from ten sources. The number of bursts
recorded in each source is given in Table~1. All bursts were of Type~1. We
refer to the review of Lewin et al. (1993) for a description of the detailed
burst properties of these sources, and for further references.

\section{Method}

In analyzing the data of a given burst source, we used the following
procedure.  First, we selected about 200~s of data around each burst. We then
subdivided each 200-s section into segments with a duration of 8~s, and kept
only those 8-s segments that had an average count rate of more than $\sim$
20\% above the persistent flux. On average this resulted in about four
accepted 8-s segments per burst, but this number varied strongly among
different bursters and between bursts. Then we calculated the power spectrum
of each selected 8-s segment using an FFT. The frequency resolution was
uniformly 0.125~Hz in all power spectra. Subsequently, all obtained power
spectra of a given source were normalized according to Leahy et al. (1983),
and summed. Finally, the average power spectrum was calculated by dividing by
each summed power by the number $M$ of summed power spectra.

During the EXOSAT observations many different time resolutions were used.
Therefore, power spectra obtained from different bursts of the same source
generally had different Nyquist frequencies, and the averaged power spectrum
had adjacent frequency ranges with different values of $M$. The resulting
different sensitivities in these frequency ranges were properly taken into
account in the analysis.

For pure Poisson noise, the powers $P_i$ in the summed power spectrum (the
``noise powers'') are $\chi^2$ distributed with $2M$ degrees of freedom. By
increasing $M$, the sensitivity is increased. As an example, for the source
4U\,1636$-$53, where EXOSAT recorded 60 bursts, the pulse amplitude
sensitivity is increased by a factor \about3.5 as compared to searching a
single burst.

The trends that are present in each 8-s data segment due to the burst
envelopes produce power at low frequency in the power spectra. It is possible
to reduce this power by fitting the trends with polynomials and subtracting
them from the light curves before doing the transforms. However, comparisons
between power spectra of raw and detrended time series showed that this
correction was not necessary: the extra power was confined to frequencies
$\approxlt 1$~Hz. An exception was the case of the source GX\,354$-$0, which
has very short bursts, so that the extra power extended to higher frequencies
(\about4~Hz). This made the sensitivity to pulsations at frequencies below
4~Hz slightly lower.

We looked for significant excesses in the averaged power spectra using our
knowledge of the Poisson noise power distribution in each frequency range and
taking into account the number of trials. When no significant excesses were
found, we estimated upper limits on the power that could have been present due
to pulsations (the ``signal power''). These upper limits were estimated from
the highest actually observed power in each frequency range (van der Klis
1989), and under the assumption that the distribution of powers from the
combined noise and signal time series is that described by Groth (1975). We
refer to Vaughan et al. (1994) for a detailed description of this method.

Finally, the upper limit on the signal power $P_{UL}$ was converted into an
upper limit on the fractional pulse amplitude $A$ for an assumed sinusoidal
pulse profile using (Leahy et al.\ \cite{Leahy et al 1983}) $A = (2P_{\rm
UL}/0.773N_{\rm ph})^{0.5} (\pi\nu_{\rm j}/2\nu_{\rm Nyq}) (\sin(\pi\nu_{\rm
j}/2\nu_{\rm Nyq}))^{-1}$, where $N_{\rm ph}$ the number of counts per
transform, ${\nu}_{\rm Nyq}$ the Nyquist frequency, and $\nu_{\rm j}$ the
frequency of the j-th bin in the power spectrum. For $N_{\rm ph}$ we used
simply the average number of counts over all data segments whose power spectra
were averaged. For $\nu_{\rm Nyq}$ and $\nu_{\rm j}$ we chose conservative
values: the maximum frequency and the central frequency of the range involved.

Our procedure of using a straight average of the individual power spectra, and
of estimating the pulse amplitude using Eq.~(1) as described above ignores the
fact that through a burst, and between bursts, the X-ray flux varies
considerably. However, in the absence of any knowledge of how the pulse
amplitude depends on the X-ray flux (or the burst phase) it is not useful to
attempt to optimize the averaging procedure or improve the amplitude
calculation beyond our simple approach.

In contrast to the case of persistent flux searches, which ideally use very
high frequency resolutions in our case the frequency resolution is low, so
that there is usually no danger that pulsation peaks will move over several
frequency bins due to orbital Doppler shifts. The maximum frequency shift due
to orbital motion $\Delta\nu_{\rm max}$ is (Vaughan et al.\ \cite{Brian
search}):

\begin{equation}
{\Delta\nu_{\rm max}\over \nu_{\rm pulse}} = 0.0021 {M_2\over M_{\rm tot}}
\left(P_{\rm orb}\over 1\,{\rm hr}\right)^{-1/3} \left(M_{\rm tot}\over
M_\odot\right)^{1/3}
\end{equation}

\noindent where $\nu_{\rm pulse}$ is the pulse frequency, $M_2$ the mass of
the donor star, $M_{\rm tot}$ the total mass of the system and $P_{\rm orb}$
the orbital period.

For a low mass X-ray binary with a 1.4~\msun\ neutron star and a Roche lobe
filling 0.1--0.3~\msun\ main sequence red dwarf companion we find that the
critical pulse frequency above which Doppler shifts can cause the pulse peak
to move over one frequency bin is between 400 and 800~Hz. For 4U\,1820$-$30,
the binary with the shortest known orbital period (685 seconds) this critical
frequency is \mbox{173 Hz} (using $M_{2}=0.28$; van der Klis et al.\
\cite{1820-30Michiel}). Only in a few cases do the Nyquist frequencies in our
search (Table~1) exceed the relevant critical frequency; in these cases,
depending on the true orbital parameters, the orbital inclination and the
orbital phases during the observations, the pulse amplitude sensitivity may
have been lowered by a factor $\sim(\nu_{\rm pulse}/\nu_{\rm crit})^{0.5}$.

The bursts whose power spectra we averaged were collected over a 3-year
period.  Therefore we also consider accretion induced spin-up or spin-down as
a cause of pulse frequency shifts. Using standard formulae (Rappaport \& Joss
\cite{Rappaport&Joss 1977}) and assuming a neutron star magnetic field
strength of $\approxlt 10^9$~G and accretion rates at or below the Eddington
critical rate, we estimate that pulse frequency shifts would amount to the
width of one frequency bin after several thousand years, and are therefore
negligible.

\section{Results and discussion}

We found no significant excesses in the average power spectra. The 99\%
confidence upper limits on the sinusoidal pulse amplitudes are listed in Table
\ref{tab:upper limits}. These upper limits are given as a fraction of the
average {\it burst} flux; they have been corrected for the persistent flux and
background. For most of our sources we obtained upper limits that are below
previously reported detections in other, but similar, sources (Aql\,X-1,
Schoelkopf and Kelley 1991, 10\%; 4U\,1608$-$52, Murakami et al. 1991,
8--20\%).

\begin{table}[ht]
\begin{center}
\small
\caption[]{Upper limits on sinusoidal amplitude (99\% confidence)}
\label{tab:upper limits}
\begin{tabular}{cccc}
\hline
            Source& Frequency range&           Number& Upper limit\\
                  &            (Hz)&        of bursts&        (\%)\\
\hline
4U\,1636$-$53 $^1$&           1--16&               60&         1.9\\
                  &          16--64&               45&         2.7\\
                  &         64--256&               12&         4.5\\
           Ser~X-1&           1--16&                3&          11\\
                  &          16--64&                1&          13\\
   EXO\,1747$-$214&           1--16&                2&          11\\
            GC X-1&           1--16&                2&     8.8$^2$\\
     4U\,1705$-$44&           1--64&               24&         3.9\\
                  &         64--256&               16&         5.9\\
                  &        256--512&               14&         7.3\\
                  &       512--2048&                6&         7.0\\
 EXO \,0748$-$676 &           1--16&               33&         4.5\\
                  &         16--128&               12&         6.6\\
                  &        128--256&                2&          18\\
  4U~1820$-$30$^3$&           1--64&                7&         3.4\\
                  &         64--256&                3&         4.8\\
     GX\,354$-$0  &           4--16&                8&         6.3\\
                  &          16--64&                5&         6.7\\
                  &          64--73&                3&         6.2\\
                  &         73--256&                1&         8.3\\
   Rapid Burster  &           1--64&                2&         8.2\\
   4U\,1735$-$44  &           1--32&                6&         9.6\\
                  &          32--64&                5&          11\\
                  &         64--256&                1&          15\\
\hline
\end{tabular}

\begin{tabular}{p{8.5cm}}

\vspace{0.5cm}

$^1$ The upper limits for 4U\,1636$-$53 were calculated from the average power
spectra of all bursts, radius expansion bursts included. The average power
spectra with these bursts omitted have also been checked. In addition the
average power spectra of all radius expansion bursts have also been
investigated for possible oscilations of the expanding atmosphere (Murakami \&
Inoue \cite{Murakami1987}).

\vspace{0.5cm}

$^2$ Only the two bigger bursts observed were used to determine the upper
limit. The two smaller bursts were also checked, but we suspect that these
came from another source.

\vspace{0.5cm}

$^3$ All bursts show radius expansion; see also note 1.

\end{tabular}
\end{center}
\end{table}

The negative result of our search has two possible interpretations: either
coherent pulsations are present in bursts, but in the sources we studied their
frequencies are higher than a few 10$^2$~Hz or their average amplitudes are
smaller than a few percent, or {\it no} coherent pulsations are normally
present in bursts. This latter possibility might apply if the nuclear burning
does not lead to anisotropic emission after all. Another possibility is that
the anisotropy is changing on time scales similar to the spin period. If the
pulse profile changes that rapidly, the signature of the rotation in the power
spectra will not be a sharp peak.

There is a number of ways in which one might improve the sensitivity of
searches such as ours. By correcting for binary orbital acceleration during
the burst, sensitivity to particularly $\gg$100~Hz pulsations could be
improved, especially in long ($\approxgt$10~s) bursts occurring in systems
with tight binary orbits. However, as the binary orbits are not known one
would need to perform a search in acceleration space (see, e.g., Vaughan et
al. 1994), and because the reconstructed pulse frequencies would still be
Doppler-shifted by different amounts in different bursts, the pulsation peaks
would end up in different frequency bins in the power spectra. {\it Coherent}
transformation of the bursts is possible in principle, but with standard
methods in practice computationally prohibitive. Collecting more bursts and
applying the method of incoherent addition that we used here will improve
pulse amplitude sensitivity as roughly the fourth root of the number of
bursts. The most likely way to success, however, is to increase detector area,
as pulse amplitude sensitivity scales as the square root of the count
rate. Searches using archival Ginga data, and with NASA's X-ray Timing
Explorer seem most promising.

\end{document}